\documentclass[a4paper, unpublished]{quantumarticle}
\pdfoutput=1

\usepackage{amsmath}
\usepackage{amssymb}
\usepackage{graphicx}
\usepackage{dcolumn}
\usepackage{bm}
\usepackage{xcolor}
\usepackage{tikz}
\usepackage{xfrac}
\usepackage{blindtext}
\usepackage{multirow}
\usepackage{chemformula}
\usepackage{braket}
\usepackage{tabularx}
\usepackage[normalem]{ulem}
\usepackage[caption=false]{subfig}

\usepackage[numbers,sort&compress]{natbib}

\definecolor{darkblue}{HTML}{004D6B}
\definecolor{darkred}{HTML}{8c1515}
\definecolor{darkgreen}{HTML}{006400}

\usepackage[
  colorlinks=true,
  urlcolor=darkred,
  citecolor=darkblue,
  linkcolor=darkred,
  breaklinks
]{hyperref}

\graphicspath{{figures/}}

\usepackage[caption=false]{subfig}

\graphicspath{{figures/}}

\newcommand{\be}{\begin{equation}}
\newcommand{\ee}{\end{equation}}

\newcommand{\bea}{\begin{eqnarray}}
\newcommand{\eea}{\end{eqnarray}}
\newcommand{\beal}{\begin{align}}
\newcommand{\eeal}{\end{align}}

\renewcommand{\vec}[1]{\ensuremath{\mathbf{#1}}}

\newcommand{\abs}[1]{\left| #1 \right|} 

\begin{document}

\title{\shortstack[l]{Floquet implementation of a 3d fermionic toric code\\
with full logical code space}}
\author{Yoshito Watanabe}
\author{Bianca Bannenberg}
\author{Simon Trebst}
\affiliation{Institute for Theoretical Physics, University of Cologne, 50937 Cologne, Germany}

\begin{abstract}
Floquet quantum error-correcting codes provide an operationally economical route to fault tolerance by dynamically generating stabilizer structures using only two-body Pauli measurements. 
But while it is well established that stabilizer codes in higher spatial dimensions gain additional levels of intrinsic robustness (allowing for larger sets of fault-tolerant logical operations and eliminating the need for active error correction), higher-dimensional Floquet codes have hitherto been explored only in limited scope.
Here we introduce a three-dimensional generalization of a Floquet code whose instantaneous stabilizer group realizes a {\it 3$d$ fermionic toric code}, while crucially preserving all three logical qubits throughout the entire measurement sequence.  
One central ingredient is the identification of a three-dimensional lattice geometry that generalizes the features of the Kekul\'e lattice underlying the two-dimensional Hastings-Haah code -- specifically, a structure where deleting \emph{any} one edge color yields a two-color subgraph that decomposes into short, closed loops rather than homologically nontrivial chains.
This loop property avoids the collapse of logical information that plagues naive sequential two-color measurement schedules on many 3$d$ lattices. 
Although, for our lattice geometry, a simple 3-round ($z \rightarrow x \rightarrow y$) cycle that sequentially measures the three parity checks / bond types  does not expose the full error syndrome set, we show that one can append an additional measurement sequence to extract the missing syndromes without disturbing the logical subspace. This yields a 10-round cycle in total.
Beyond code design, 3$d$ tricoordinated lattice geometries define a family of 3$d$ {\it monitored Kitaev models}, in which random measurements of the non-commuting parity checks give rise to dynamically created entangled phases with nontrivial topology.
In discussing the general structure of their underlying phase diagrams and, in particular, the existence of certain quantum critical points, we again make a connection to the general preservation of logical information in time-ordered Floquet protocols.
\end{abstract}

\maketitle
\hypersetup{
  pdftitle={Floquet implementation of a 3d fermionic toric code with full logical code space}
}

\section{Introduction}

In the quest to build a fault-tolerant quantum computer, there is an ongoing effort to design quantum error-correcting codes that can be implemented with minimal physical and operational overhead~\cite{Terhal2015}. One of the most studied codes is the toric code (TC)~\cite{Dennis2002, Kitaev2003}, an instance of a stabilizer code which is typically realized in its planar variant, the surface code~\cite{Fowler2012}, as in recent experimental work~\cite{Acharya2025} that has demonstrated error-correcting performance below threshold.  
In parallel to these experimental developments, there has been a recent surge of interest in {\it dynamically generated} quantum codes, so-called Floquet codes~\cite{Hastings2021, Haah2022, Aasen2022}, which promise a much smaller operational overhead than the four-body measurements typical of syndrome extraction in the TC by employing a periodic sequence of two-body Pauli measurements. The key idea is to design a measurement sequence so that the instantaneous stabilizer group (ISG) at certain times is equivalent to that of a stabilizer code, thereby allowing one to encode quantum information in the logical subspace of the code. 
A paradigmatic example is the Floquet code introduced by Hastings and Haah~\cite{Hastings2021, Haah2022}, which employs a Kekul\'e-type edge coloring of the honeycomb lattice~\cite{Schmidt10honeycomb} and a periodic sequence of mutually non-commuting two-qubit measurements (e.g., identifying the three colors with $XX$, $YY$, and $ZZ$ parity checks, respectively). This Floquet code realizes the TC ISG at each time step of its measurement sequence, thereby enabling the encoding of logical qubits with error-correcting capabilities~\cite{Gidney2021, Gidney2022}. 

\begin{figure*}
    \centering
    \includegraphics[width=1.0\textwidth]{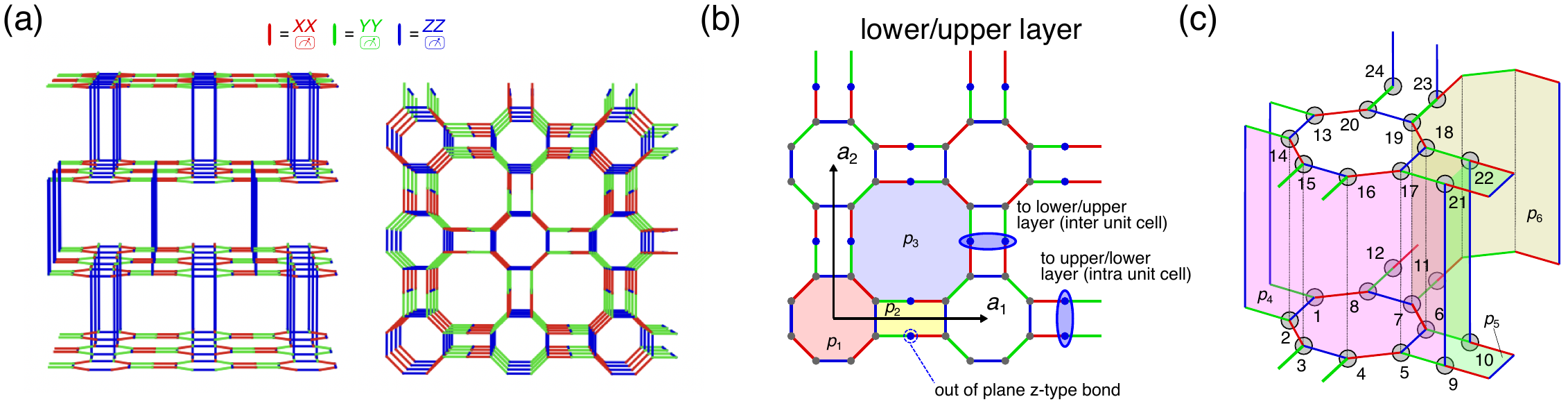}
    \caption{
    {\bf Three-dimensional Kekul\'e-Kitaev lattice.} 
     The lattice is a 3$d$ tricoordinated lattice that allows a three-edge coloring.
    (a) Two perspectives of the lattice: square--octagon layers are stacked along $\vec a_3$ and coupled by vertical $z$-type (blue) bonds.
    (b) Projection of one bilayer unit cell onto the $\vec a_1$--$\vec a_2$ plane. Each layer is based on a square--octagon lattice in which every square is decorated by additional vertices that enable three-dimensional connectivity; the unit cell contains a lower and an upper layer.
    The vertical ($\parallel\vec a_3$) $z$-type bonds have two inequivalent in-plane footprints within the unit cell, half-shifted along $\vec a_1$ or $\vec a_2$, corresponding to intra- and inter-unit-cell interlayer couplings (blue ovals).
    (c) Examples of elementary non-coplanar plaquettes involving out-of-plane bonds; plaquettes of types $p_4$, $p_5$, and $p_6$ are shown. 
    Ignoring the edge coloring, the in-plane primitive translations are $\vec a_1$ and $\vec a_2$; each layer has a 12-site unit cell (24 sites per bilayer). The edge coloring doubles the translation period along both $\vec a_1$ and $\vec a_2$, giving 96 sites in the full (colored) unit cell (a $45^\circ$-rotated choice of lattice vectors reduces this to 64 sites, but we use the unrotated convention). We label system size by $(L_1,L_2,L_3)$, the number of unit cells along $(\vec a_1,\vec a_2,\vec a_3)$; a consistent coloring under periodic boundary conditions requires $L_1$ and $L_2$ to be even.}
    \label{fig:lattice}
\end{figure*}

It is natural to ask how far such constructions can be extended to three spatial dimensions~\cite{Zhang2022,Davydova2023, Davydova2024, Xu2026}. From a conceptual viewpoint, it has long been appreciated that higher spatial dimensions lead to thermal stability of stabilizer codes, manifest themselves in finite-temperature transitions \cite{Castelnovo2008}, and stable quantum memories, e.g.\ for the four-dimensional toric code \cite{Alicki2010}. For the physically realizable case of three spatial dimensions, it has recently been shown that the {\it 3$d$ fermionic toric code} (fTC)~\cite{Levin2003, Keyserlingk2015, Chen2019} can sustain topological order and long-range entanglement (though not a logical qubit) at finite temperatures \cite{Zhou2025}.
From the quantum error-correction viewpoint, 3$d$ topological and subsystem codes are attractive because their higher-dimensional locality can support a richer set of locality-preserving logical gates than in two spatial dimensions (2$d$)~\cite{Bravyi2013, Pastawski2015, Vasmer2019}, enabling routes to universality via fault-tolerant code conversion (code switching) between codes with complementary protected gate sets~\cite{Anderson2014, Bombin2016}. Moreover, some 3$d$ architectures exhibit sufficient redundancy among local checks to enable single-shot syndrome extraction~\cite{Bombin2015, Brown2016, Kubica2022}.

In this work, we explore a three-dimensional generalization of dynamically generated quantum codes, in particular, a Floquet implementation of the 3$d$ fTC. 
Our starting point is a 3$d$ generalization of the honeycomb Kitaev model \cite{Kitaev2006} to the lattice structure shown in Fig.~\ref{fig:lattice}
whose tricoordinated geometry allows one to extend a recurring motif of 2$d$ Floquet codes -- sequences of non-commuting two-body parity-checks -- to a setting with genuine error-correcting capability in 3$d$.
In doing so, our work follows in the footsteps of earlier work~\cite{Dua2024}, where a 3$d$ fTC was constructed from the 3$d$ Kitaev model on the hyperhoneycomb lattice~\cite{Mandal2009,OBrien2016}, dynamically generating a single logical qubit over a 16-round measurement cycle.
Our lattice and protocol likewise realize a 3$d$ fTC as an ISG, but crucially preserves \emph{all three} logical qubits throughout the Floquet measurement sequence. A key design criterion is the structure of the two-color subgraphs obtained by \emph{deleting all edges of any one color}. For many known lattices~\cite{Mandal2009, Hermanns2014, Nasu2014, Nasu2014_2, Hermanns2015, Hermanns2015_2, OBrien2016, Eschmann2020}, removing a given color/bond type leaves behind homologically nontrivial chains (non-contractible loops); these chains are, however, directly associated with logical operators, making simple successive two-color measurement schedules prone to collapsing the logical information. Our lattice avoids this failure mode by ensuring that, upon removing \emph{any} one color/bond type, the remaining subgraph decomposes into a collection of short, closed loops -- objects that are distinct from the logical operators. While the naive $z \rightarrow x \rightarrow y$ Floquet sequence alone does not provide access to all error syndromes, we show that one can append an additional measurement schedule to extract the missing syndromes without disturbing the logical information, resulting in a 10-round measurement cycle in total. In our case, one full cycle implements a trivial automorphism.

Beyond their role in error correction, the same ingredients that underlie Floquet codes -- local, non-commuting two-qubit Pauli measurements on a $3$-edge-colored tricoordinated lattice -- also allow one to define {\it monitored Kitaev models}~\cite{Lavasani2023, Sriram2023,Zhu2024,Klocke2025}, a family of models whose measurement-only (monitored) many-body dynamics~\cite{Lang2020,Ippoliti2021} gives rise to steady-state ensembles of entangled states of matter with nontrivial topology. In this setting, one repeatedly measures  randomly chosen bond operators, and the relative rates at which the three bond types are randomly selected provide simple control parameters, analogous to tuning couplings in the Hamiltonian Kitaev model. Varying these rates leads to distinct entanglement-scaling regimes and motivates a ``measurement phase diagram'' for a given lattice geometry that we argue to provide important guidance also for the setup of a Floquet code. 
Based on Clifford-circuit simulations on several three-dimensional tricoordinated lattice geometries (beyond the Kekul\'e lattice structure at the heart of this manuscript),
we argue that Hamiltonian gapless regimes typically broaden into measurement-induced critical phases, often extending beyond their Hamiltonian counterparts. In addition, we tie the presence or absence of certain quantum critical points in these phase diagrams, protected by a geometric property, 
to the preservation of logical information under time-ordered Floquet sequences. \\

The remainder of the paper is organized as follows. In Sec.~\ref{sec:floquet_code} we first review the key ideas behind 2$d$ Floquet codes on the (Kekul\'e) honeycomb lattice and then introduce our 3$d$ generalization, including the underlying tricoordinated lattice geometry, the instantaneous stabilizer structure, and the measurement schedule that yields a 10-round cycle while preserving the full logical subspace. We then analyze in detail how error syndromes are extracted along the cycle, how the logical operators are represented and protected against inadvertent measurement. Finally, in Sec.~\ref{sec:random_measurement_dynamics}, we discuss the broader connection to monitored Kitaev dynamics on 3$d$ tricoordinated lattices and present numerical evidence for the resulting measurement-induced phase structure, before concluding with a discussion of implications for fault-tolerant logical control and open problems in Sec.~\ref{sec:discussion}.

\begin{figure}[b]
    \centering
    \includegraphics[width=1.0\columnwidth]{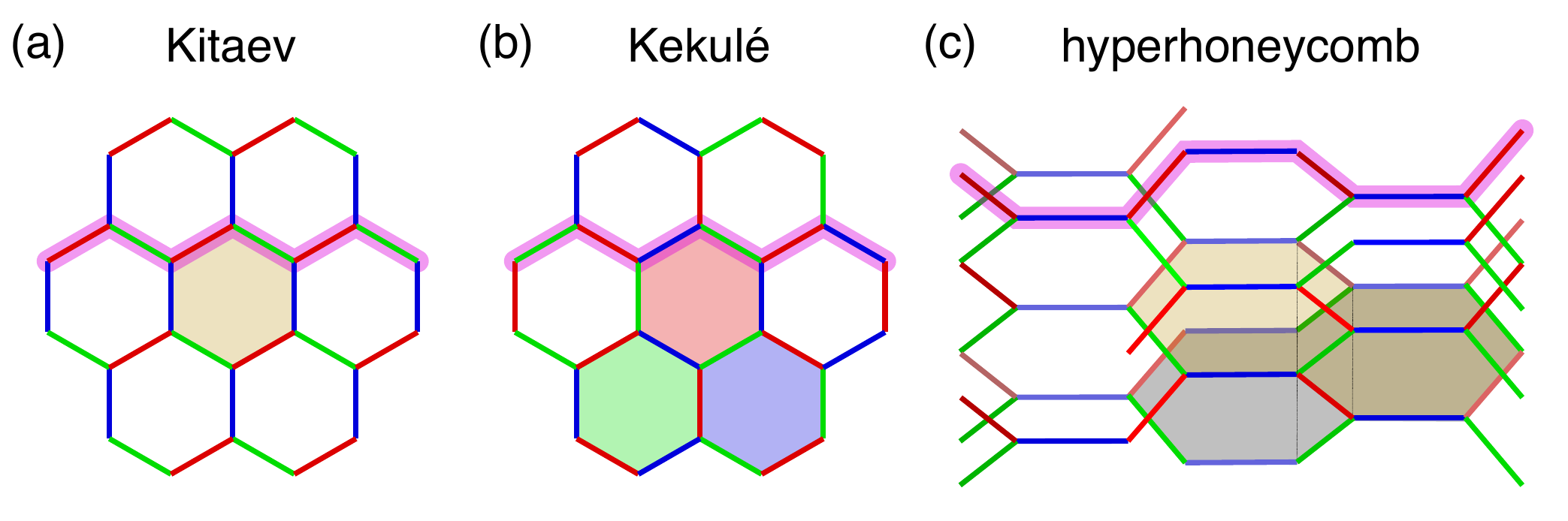}
    \caption{{\bf Tricoordinated lattice geometries.}
    (a) Honeycomb lattice with Kitaev edge coloring. 
    (b) Honeycomb lattice with Kekul\'e edge coloring. 
    (c) Hyperhoneycomb lattice. 
    In (a) and (c), the shaded plaquettes are three-colored (their boundaries contain all three edge colors). In (b), the plaquettes are two-colored; we label each plaquette by the \emph{missing} color (indicated by the shading), e.g., an $x$-plaquette is bounded by $y$- and $z$-colored edges. 
    Examples of inner logical operators are shown as thick pink lines in each panel.}
    \label{fig:lattice_2}
\end{figure}

\section{Floquet code}
\label{sec:floquet_code}

The essential ingredient of the original Hastings-Haah code is a dynamical implementation of the Kekul\'e variant  of the honeycomb Kitaev model, which, to make this manuscript self-contained, we will briefly review in the following. We will then turn to our 3$d$ code, which is based on a 3$d$  generalization of the Kekul\'e-Kitaev lattice, whose genesis and features we will introduce to discuss more general design principles for 3$d$ Floquet codes.

\subsection{Review of 2$\mathbf{d}$ Floquet codes}
A Floquet code is a dynamically generated stabilizer code obtained from a \emph{properly} designed periodic sequence of (typically non-commuting) Pauli check measurements $\mathcal{O}=\{O_i\}\subset\mathcal{P}_n$, where $\mathcal{P}_n$ is the $n$-qubit Pauli group. Given a measurement record, the post-measurement state is stabilized by an instantaneous stabilizer group (ISG) $\mathcal{S}(t)=\langle S_j(t)\rangle$, whose generators $S_j(t)\in\mathcal{P}_n$ are mutually commuting and independent, with $-I\notin\mathcal{S}(t)$~\footnote{We distinguish the (time-independent) \emph{code} stabilizer group $\mathcal{S}$, which defines a codespace, from the \emph{instantaneous} stabilizer group $\mathcal{S}(t)$, which stabilizes the post-measurement state and depends on measurement outcomes.}.  The schedule is engineered so that, at selected times, the ISG is equivalent to that of a target stabilizer code (possibly up to a Clifford frame), allowing one to store logical information in its logical subspace. It is often convenient to interpret the measured checks as generators of a subsystem-code gauge group $\mathcal{G}=\langle O_i\rangle$~\cite{Poulin2005,Bacon2006,Aliferis2007,Bravyi2011,Bombin2010}.

As a canonical example, consider the honeycomb lattice with edges colored in the usual Kitaev fashion~\cite{Kitaev2006}, as shown in Fig.~\ref{fig:lattice_2}(a). In the anisotropic coupling regimes it realizes TC topological order, with plaquette fluxes $W_p$ acting as stabilizers; their eigenvalues (syndromes) can be inferred from two-body parity-check measurements. When interpreted as a subsystem code this construction has \emph{no} logical qubits~\cite{Suchara2011}, since the non-contractible loop operators that would serve as logical operators lie in the gauge group (the \emph{inner} logical operators). Consequently, one must carefully design the measurement sequence to avoid \emph{inadvertently} measuring such logical operators; indeed, simply measuring all $x$, $y$, and $z$ edges of the conventional Kitaev honeycomb coloring in succession will measure logical operators and thereby annihilate the encoded logical subspace.

This issue can be avoided by instead using a Kekul\'e coloring of the honeycomb lattice [Fig.~\ref{fig:lattice_2}(b)]. With this coloring, measuring the three colors in succession (which we identify with $z$, $x$, and $y$ bonds, respectively, although in principle the schedule and the Pauli type associated with each edge can be chosen independently), one obtains a Floquet code that preserves the logical subspace~\cite{Hastings2021}. The key geometric point is that removing all edges of any one of the three colors leaves only disconnected \emph{finite cycles}. As a result, operators generated within two successive color rounds are contractible and cannot coincide with a non-contractible inner-logical string.
Another useful property of this Kekul\'e coloring is that each plaquette involves only two of the three edge colors. This seemingly minor constraint is crucial: it allows one to extract the full set of plaquette syndromes using only checks from adjacent color rounds, and it greatly simplifies the scheduling logic -- the syndrome-extraction structure is essentially dictated by the edge coloring alone.

\begin{figure*}
    \includegraphics[width=1.0\textwidth]{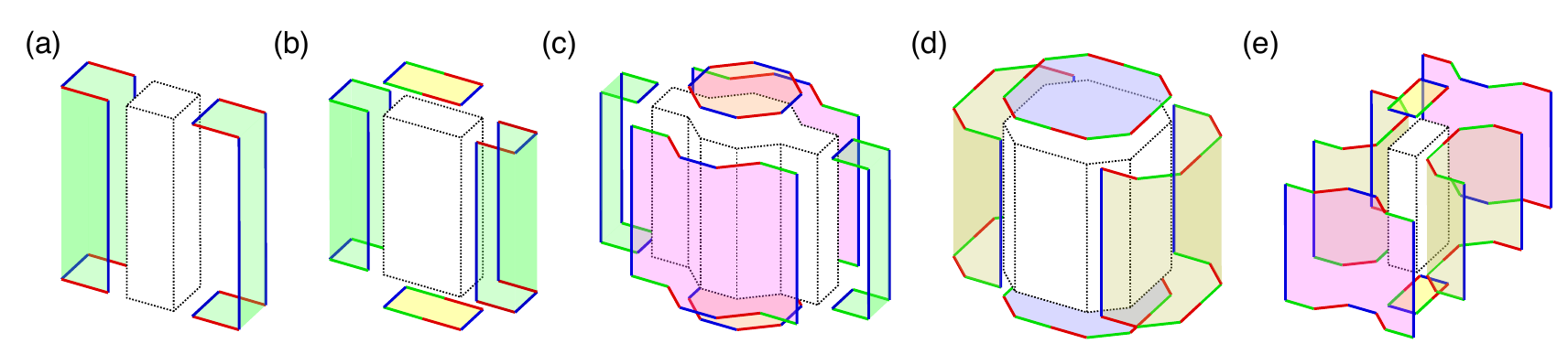}
    \caption{{\bf Enclosed volumes.} 
    A volume is formed by a set of plaquettes (shaded) whose union is a closed surface enclosing a three-dimensional region. The product of the plaquette operators on the boundary of such a volume equals the identity, yielding a volume constraint. 
    (a) A $p_5$ plaquette can be replaced by an alternative choice of surface spanning the same boundary loop: the two surface choices are equivalent because together they form a closed surface that encloses a volume. 
    (b--e) Other examples of enclosed volumes formed from combinations of plaquette types.}
    \label{fig:volumes}
\end{figure*}

\subsection{Generalization to 3$\mathbf{d}$ Floquet codes}
Our goal here is to generalize these ideas to three spatial dimensions. To formalize the relevant properties, we seek a periodic three-dimensional graph $G$ satisfying:
\begin{itemize}
 \item[(1)] $G$ is tricoordinated and admits a Kitaev-type $3$-edge-coloring by $\{x,y,z\}$ such that no two edges of the same color meet at a vertex. 
 \item[(2)] Removing any single color (bond type) yields subgraphs that are only finite cycles and, in particular, no bi-infinite chains (homologically nontrivial).
 		That is, for a given color, say $z$, one has to consider the subgraph $G_{xy}$ obtained by deleting all $z$-colored edges. 
 		Then $G_{xy}$ is $2$-regular and hence a disjoint union of cycles. 
		We impose the condition that \emph{every} connected component of $G_{xy}$ is finite in the infinite periodic system. 
 \item[(3)] The embedding of $G$ admits a well-defined set of elementary plaquettes, which serves as a syndrome check, 
 and each plaquette uses \emph{exactly two} of the three edge colors. 
\end{itemize}

The standard honeycomb lattice admits both the Kitaev and Kekul\'e variants, and both satisfy condition (1). However, the Kitaev coloring uses all three colors around each hexagonal plaquette, and removing any one color produces bi-infinite chains, meaning that conditions (2)--(3) are violated. The Kekul\'e variant, on the other hand, produces only two-colored plaquettes, and removing any single color yields only finite loops. Thus, the desired structure \emph{can} be realized in 2$d$ (see Appendix~\ref{app:2d_models}).

The central question in going to a 3$d$ generalization then is whether a ``3$d$ Kekul\'e--Kitaev'' structure exists. While a vast family of 3$d$ tricoordinated lattice geometries has been studied in the context of 3$d$ Kitaev models \cite{Mandal2009, Hermanns2014, Nasu2014, Nasu2014_2, Hermanns2015, Hermanns2015_2, OBrien2016, Eschmann2020}, none of these three-colored lattices satisfies all of these conditions, and it remains unclear whether the conditions (1)--(3) genuinely conflict in three dimensions. In attempting to realize such a lattice, a recurring obstacle is that, in 3$d$ tricoordinated lattices, elementary plaquettes can share an extended boundary segment rather than a single edge as, for instance, in the hyperhoneycomb lattice, Fig.~\ref{fig:lattice_2}(c). This leads to the following constraint: if one plaquette uses only two colors, then an adjacent plaquette that overlaps along a boundary segment must use the same two colors in order to satisfy condition (3). However, a three-edge-colored 3$d$ tricoordinated lattice cannot be built solely from two-color plaquettes.

One might thus ask what type of relaxations of conditions (1)--(3) are possible and at what cost. For instance, one might consider relaxing condition (3), in allowing for {\it some} plaquettes to contain all three colors -- a step that might result in additional effort for syndrome extraction. This is what we will pursue in our construction below. Alternatively, one might consider relaxing condition (2), e.g.\ in the form of giving up the ``finite-loop'' property for {\it some} colors 
-- a step that might result in losing some of the logical code space. 
This has been playing out in the hyperhoneycomb example of Ref.~\cite{Dua2024}, where none of the three colors satisfies the finite-loop property; accordingly, the Floquet sequence must be specified using additional labels beyond the edge coloring, and the resulting protocol preserves only a single logical qubit.

\subsubsection*{Lattice construction --  3$d$ Kekul\'e-Kitaev lattice}
We have constructed a lattice and coloring that satisfy the relaxed conditions described above -- namely, conditions (1) and (2), while  condition (3) does not hold for a subset of elementary plaquettes. 
This explicit example, which we have dubbed the 3$d$ Kekul\'e-Kitaev lattice, is shown in Fig.~\ref{fig:lattice}, where representatives of all elementary plaquette types are also indicated.

Ignoring the edge coloring, the in-plane primitive translations are $\vec a_1$ and $\vec a_2$; each layer has a 12-site unit cell (24 sites per bilayer). The out-of-plane $z$-type bonds run parallel to $\vec a_3$ and connect the lower and upper layers. In the $\vec a_1$--$\vec a_2$ projection [Fig.~\ref{fig:lattice}(b)], these interlayer bonds appear in closely spaced pairs (blue ovals). The ovals split into two families according to their in-plane offsets (along $\vec a_1$ versus along $\vec a_2$), which realize intra-unit-cell and inter-unit-cell interlayer couplings, respectively.

Elementary plaquettes fall into three coplanar types $p_1$, $p_2$, and $p_3$ [Fig.~\ref{fig:lattice}(b)] and three non-coplanar types $p_4$, $p_5$, and $p_6$ involving out-of-plane bonds [Fig.~\ref{fig:lattice}(c)]. Altogether, the (uncolored) unit cell contains 20 elementary plaquettes: two each of types $p_1$ and $p_3$, and four each of types $p_2$, $p_4$, $p_5$, and $p_6$. These plaquette operators are not all independent because of volume constraints -- the product of plaquette operators over any closed surface equals the identity. Four such enclosed volumes at intra-unit-cell positions are shown in Fig.~\ref{fig:volumes}(b--e); the remaining four, at inter-unit-cell positions, are obtained by the screw operation (a translation by $\vec a_3/2$ combined with a $90^\circ$ rotation about the $\vec a_3$ axis). Accounting for these constraints leaves 12 linearly independent plaquette operators per unit cell.

The edge coloring doubles the translation period along both $\vec a_1$ and $\vec a_2$, giving 96 sites in the full (colored) unit cell (a $45^\circ$-rotated choice of lattice vectors reduces this to 64 sites, but we use the unrotated convention). We label the system size by $(L_1,L_2,L_3)$, representing the number of unit cells along $(\vec a_1,\vec a_2,\vec a_3)$; a consistent coloring under periodic boundary conditions requires $L_1$ and $L_2$ to be even.

Most importantly, deleting all edges of any one color decomposes the lattice into disconnected finite components; the remaining graph therefore contains no non-contractible loops. For example, removing all $z$-colored edges leaves only finite $xy$ loops, each of which bounds a $p_3$ plaquette. Removing all $x$-colored edges leaves only finite loops built from the remaining two colors -- namely the boundaries of $p_1$ and $p_5$ plaquettes (together with their $\vec a_3$-screw images) -- and these occur only in unit cells labeled by $n\vec a_1+m\vec a_2+l\vec a_3$ with $n+m$ even. Removing all $y$-colored edges is analogous, except that the corresponding $p_1$ and $p_5$ plaquettes occur in unit cells with $n+m$ odd.

\begin{figure}[t]
    \includegraphics[width=1.0\columnwidth]{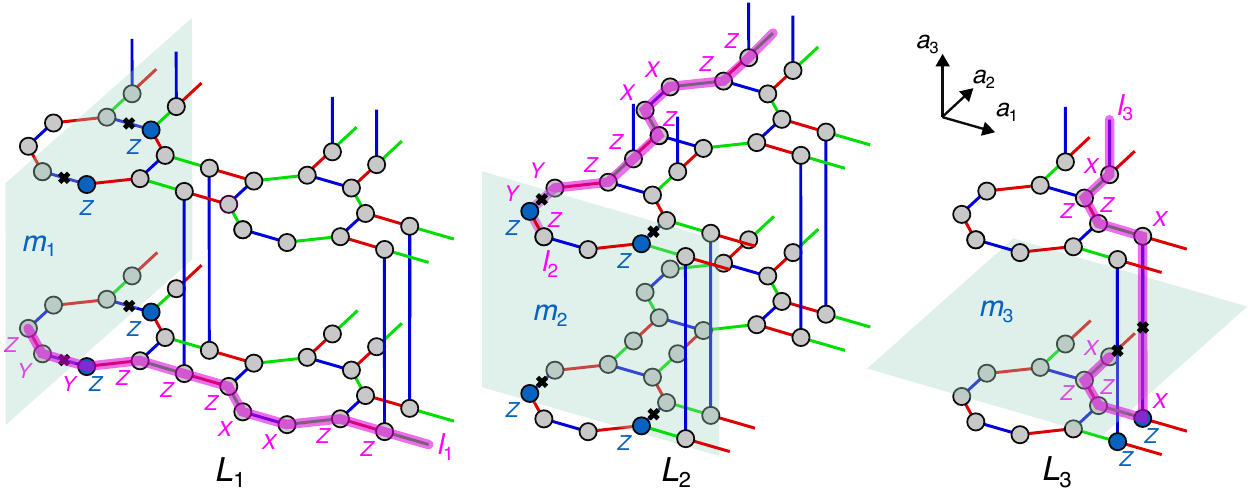}
    \caption{{\bf Logical Pauli operators for the three encoded qubits.}
    For each logical qubit $L_i$ ($i=1,2,3$), we choose a pair of non-contractible operators: a line operator $l_i$ running along the $\vec a_i$ direction (purple) and a membrane operator $m_i$ (green) oriented transverse to $l_i$. The line operators are inner logicals: they lie in the gauge group generated by the measured checks, commute with all checks, and are therefore round-independent. The membrane operators are outer logicals and can be round-dependent; shown are representatives after a $z$ round, constructed by placing a single-site $Z$ on one endpoint of each $z$-colored bond intersected by the membrane (intersections marked by crosses).}
    \label{fig:logical_paulis}
\end{figure}

\subsubsection*{Logical operators}
Similar to other 3$d$ tricoordinated lattices with three-edge coloring, our lattice admits a fixed-point, gapped Hamiltonian in a strongly anisotropic limit, whose mutually commuting terms consist of plaquette operators together with a chosen set of two-body bond terms of a fixed color $c$. Its ground state realizes a fermionic TC phase, which we denote as fTC$_c$. When defined on a three-torus $T^3$, the model encodes three logical qubits. A convenient choice of logical Pauli operators is shown in Fig.~\ref{fig:logical_paulis}. Each logical qubit $L_i$ ($i=1,2,3$) is specified by a pair of non-contractible operators: a line operator $l_i$ running along the $\vec a_i$ direction and a membrane operator $m_i$ oriented transverse to $l_i$. The line operators (typically chosen as logical $Z$) are products of the corresponding two-body bond Pauli operators along a non-contractible line. By construction, they commute with all measured checks; equivalently, in the Floquet language, they are inner logicals and are therefore round-independent. As we will see, in the fTC phase, these line operators are Wilson loop operators of the emergent fermionic charge; they generate an anomalous 2-form symmetry in $(3+1)D$~\cite{Gaiotto2015}.

The membrane operators (typically chosen as logical $X$) are complementary outer logicals and can be round-dependent. Figure~\ref{fig:logical_paulis} shows representatives after a $z$ round (i.e., for the fTC$_z$ phase): one chooses a membrane surface transverse to $\vec a_i$ and places a single-site $Z$ on one endpoint of each $z$-colored bond intersected by the membrane (intersections are marked by crosses).  For $L_1$ and $L_2$, $x$- and $y$-round representatives are obtained by shifting the support to the next sites (along $+\vec a_1$) and replacing $Z\mapsto X$ or $Z\mapsto Y$, respectively. For $L_3$, the $x$- and $y$-round representatives follow from the $z$-round membrane by locally replacing its two-site support on each relevant $p_2$ plaquette by a four-site support on the same plaquette, with $Z\mapsto X$ or $Z\mapsto Y$.

\begin{figure}[b]
	\includegraphics[width=1.0\columnwidth]{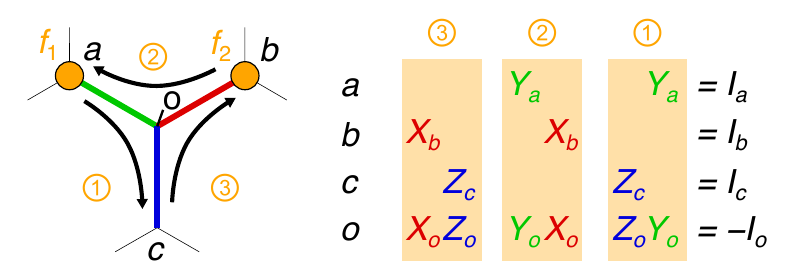}
    	\caption{{\bf  T-junction--exchange.} 
            Two point-like charges $f_1$ and $f_2$ at the endpoints $a$ and $b$ are exchanged via the three-step sequence \textcircled{\scriptsize 1}--\textcircled{\scriptsize 3} around a trivalent \((x,y,z)\) Kitaev vertex \(o\).  Each step is implemented by a short string operator given by the product of the two adjacent bond operators meeting at \(o\). The junction contribution picks up a minus sign, diagnosing fermionic self-statistics of the charge excitation (independent of spatial dimensionality).}
    \label{fig:t_junction_exchange}
\end{figure}

\subsubsection*{Fermionic nature of point-like excitations}

For the honeycomb Floquet code, it has been shown that the inner-logical operators carry {\it fermionic} charge excitations~\cite{Hastings2021}, which can be diagnosed via the T-junction--exchange algebra of string operators~\cite{Ma2023, Liu2024}. This is schematically illustrated in Fig.~\ref{fig:t_junction_exchange} below. More generally, the same algebraic argument applies to any lattice satisfying condition~(1), also in three spatial dimensions: it implies a (deconfined) point-like excitation with fermionic self-statistics, localized at the endpoints of an open string operator~\footnote{We thank Nathanan Tantivasadakarn for pointing this out.}. 
In contrast to the 2$d$ case, where a fermion can arise as a bound state of two point-like bosonic anyons (e.g., $\epsilon=e\times m$ in the toric code), in 3$d$ this fermionic point excitation is intrinsically fermionic and cannot be decomposed into two bosonic point charges; this distinguishes 3$d$ fTC from the conventional (bosonic) toric code. 
An explicit finite-depth local unitary circuit equivalence between the fTC and the $\mathbb{Z}_2$ topologically ordered phase of the hyperhoneycomb model was established in Ref.~\cite{Dua2024} using an entanglement-renormalization analysis -- with the same line of argument expected to hold also for our 3$d$ lattice geometry. The mutual statistics between point-like and loop-like excitations can be diagnosed by the same operator-algebra reasoning: a closed line operator, constructed as a product of plaquette operators along a loop that links a flux loop created by a membrane operator, acquires a linking phase that captures the nontrivial charge--loop braiding.

\subsection{Measurement schedule}
Completing our construction, we now turn to the design of a periodic measurement schedule that preserves all three logical qubits throughout the Floquet sequence. We label each measurement round by an edge color; the two-qubit Pauli operator measured on a bond is fixed by that color (as in the Kitaev-type construction). The backbone of the schedule is the color cycle $z \rightarrow x \rightarrow y \rightarrow z$, which by construction avoids directly measuring any logical operator. Whenever the cycle closes (i.e., upon returning to a $z$ round), the measurement record also determines the plaquette syndromes supported on the two colors measured in adjacent rounds; in our lattice, these are the $p_1$, $p_3$, and $p_5$ plaquettes.

Unlike the 2$d$ Kekul\'e honeycomb Floquet code, however, these do not form a complete, independent set of plaquette checks in three dimensions, necessitating additional measurement rounds to access the remaining syndromes. Specifically, we add measurements to extract all $p_2$ syndromes and to access a subset of the $p_4$ and $p_6$ syndromes; the remaining $p_4$ and $p_6$ syndromes can then be inferred using the volume constraints shown in Fig.~\ref{fig:volumes}(c--e). Note that the basic Floquet cycle already yields all $p_5$ syndromes, and we additionally measure all $p_2$ plaquettes; consequently, our syndrome record is overcomplete because of the volume constraint in Fig.~\ref{fig:volumes}(b). Such redundancy may be beneficial for error correction in the presence of measurement errors.

Concretely, the 10-round schedule is
\begin{equation}
(z,\ x,\ y,\ z_{\rm intra}^{\rm even},\ x,\ z_{p_2},\ y_{p_2},\ x,\ z_{\rm inter}^{\rm even},\ y),
\end{equation}
where $x$, $y$, and $z$ denote full-color rounds (all edges of that color). The rounds $z_{p_2}$ and $y_{p_2}$ are restricted to the edges participating in the four $p_2$ plaquettes. The selective rounds $z_{\rm intra}^{\rm even}$ and $z_{\rm inter}^{\rm even}$ include the corresponding out-of-plane $z$-edges (intra- and inter-unit-cell type, respectively) in all unit cells, together with an additional parity-selected set of in-plane $z$-edges in unit cells with $n+m$ even. In the site labeling of Fig.~\ref{fig:lattice}(c), we choose
\begin{equation}
\begin{aligned}
&z_{\rm intra}^{\rm even}=\{(Z_{3}Z_{4})^{\rm even},\ (Z_{15}Z_{16})^{\rm even},\ Z_{9}Z_{21},\ Z_{10}Z_{22}\},\\
&z_{\rm inter}^{\rm even}=\{(Z_{5}Z_{6})^{\rm even},\ (Z_{17}Z_{18})^{\rm even},\ Z_{11}Z_{23},\ Z_{12}Z_{24}\},
\end{aligned}
\end{equation}
where the superscript ``even'' indicates that the corresponding bond is included only in unit cells with $n+m$ even.

The three-round windows (3--5) and (8--10) provide the additional information needed to extract the $p_4$ and $p_6$ syndromes: (3--5) targets the intra-unit-cell plaquettes, while (8--10) targets the inter-unit-cell plaquettes. Rounds (5--8) extract the $p_2$ syndromes. After one 10-round cycle, all plaquette syndromes are available either directly from the measurement record or indirectly via the volume constraints, and the schedule then repeats from round (1). Although rounds (4), (6--7), and (9) require additional site, plaquette, and unit-cell labels beyond the color label alone, the overall construction remains largely dictated by the three-edge coloring. 

A straightforward operator-algebra calculation shows that the three-round cycle $z\rightarrow x\rightarrow y\rightarrow z$ acts trivially on the logical operators (up to multiplication by measured checks). We have also checked numerically that
$\langle \{ZZ\}, \{W_p\}, L\rangle$ is invariant for several choices of logical states $L$ (logical $Z$- or $X$-eigenstate choices) and that the full 10-round schedule also implements the trivial logical automorphism.

\subsection{3$\mathbf{d}$ Floquet code}

Taken together, the combination of the 3$d$ Kekul\'e-Kitaev lattice -- constructed following the design constraints (1)--(3) with a carefully chosen relaxation --
and an adopted measurement protocol -- expanding the standard 3-cycle with supplementary syndrome measurements to 10-cycle -- results in our proposal for 3$d$ Floquet
code whose instantaneous stabilizer group is a 3$d$ fTC. By construction, it preserves all three logical qubits over the entirety of the Floquet cycle and allows for active error correction (with a level of built-in redundancy that could be exploited in the presence of syndrome extraction errors).
We will further discuss this 3$d$ Floquet code in the discussion Section below.

\section{Random measurement dynamics}
\label{sec:random_measurement_dynamics}
\begin{figure}[h]
    \includegraphics[width=1.0\columnwidth]{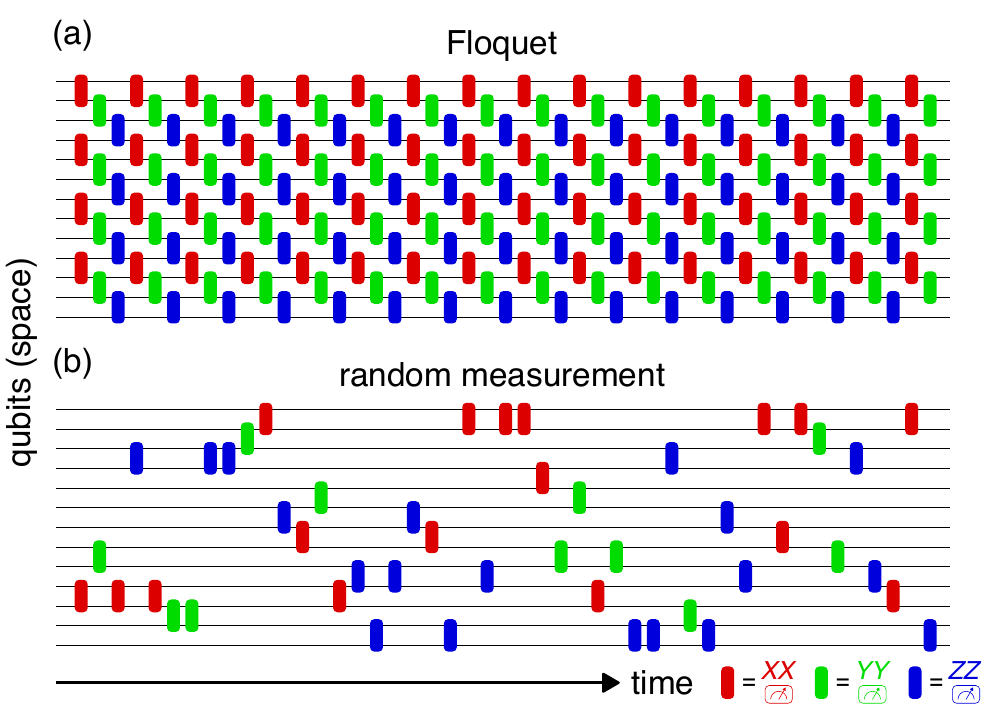}
    \caption{{\bf Floquet protocol vs.\ measurement dynamics.} (a) Floquet protocol: a periodic sequence of two-body Pauli measurements, with $XX$, $YY$, and $ZZ$ rounds shown as red, green, and blue blocks, respectively. (b) Randomized measurement dynamics: at each time step, a bond is chosen at random, and the corresponding two-body Pauli operator is measured, with the bond type sampled according to prescribed rates.}
    \label{fig:floquet_vs_md}
\end{figure}

The same structural ingredients that underlie Floquet-code constructions -- namely, local two-qubit Pauli measurements on a $3$-edge-colored tricoordinated lattice -- also define a natural class of random-measurement many-body dynamics, i.e., stochastic protocols in which the measured bonds are chosen randomly among the three colors
(see the schematic illustration in Fig.~\ref{fig:floquet_vs_md}). Our central geometric criterion from the Floquet setting -- that deleting any one color leaves only finite loops -- will reappear here as a useful organizing principle for the global structure of the measurement phase diagram in three dimensions. As a warm-up, we first review the 2$d$ monitored Kitaev honeycomb model~\cite{Lavasani2023, Sriram2023, Zhu2024, Klocke2025}, including its Kekul\'e-type variant, before turning to two representative 3$d$ tricoordinated lattices -- the hyperhoneycomb geometry put forward in Ref.~\cite{Dua2024} and the Kekul\'e-Kitaev lattice at the heart of the current manuscript.

\subsection{Review of the 2$\mathbf{d}$ monitored Kitaev model}

A convenient parametrization of the monitored dynamics assigns a measurement probability $p_c$ to each bond color $c\in\{x,y,z\}$. In each measurement step, a bond color is chosen with probabilities $(p_x,p_y,p_z)$, and a corresponding two-qubit Pauli operator on a bond of that color is projectively measured (for instance, measuring $X_iX_j$ on an $x$-bond, and similarly for $y$ and $z$). A schematic illustration of such random-measurement dynamics is provided in Fig.~\ref{fig:floquet_vs_md}. The probabilities satisfy
\begin{equation}
p_x,p_y,p_z\in[0,1],\qquad p_x+p_y+p_z=1,
\end{equation}
and therefore live on a two-dimensional simplex, often visualized as a triangle. The corners $(1,0,0)$, $(0,1,0)$, and $(0,0,1)$ correspond to dynamics dominated by measurements of a single bond color.

It is often useful to compare this measurement-only protocol with the {\it Hamiltonian} Kitaev honeycomb model by loosely identifying measurement probabilities with effective couplings, i.e., $(p_x,p_y,p_z)$ with normalized bond-directional coupling strengths $(J_x,J_y,J_z)$. While this identification is not exact -- the monitored dynamics is not generated by a Hamiltonian and depends on the measurement outcomes, i.e.\ it arises from Born probabilities and not energetics -- it provides a helpful guide for organizing the phase structure. In particular, one finds distinct steady-state entanglement regimes across the simplex, forming a ``measurement phase diagram'' in $(p_x,p_y,p_z)$ space that qualitatively echoes key features of the Hamiltonian phase diagram.

\begin{figure}
    \includegraphics[width=1.0\columnwidth]{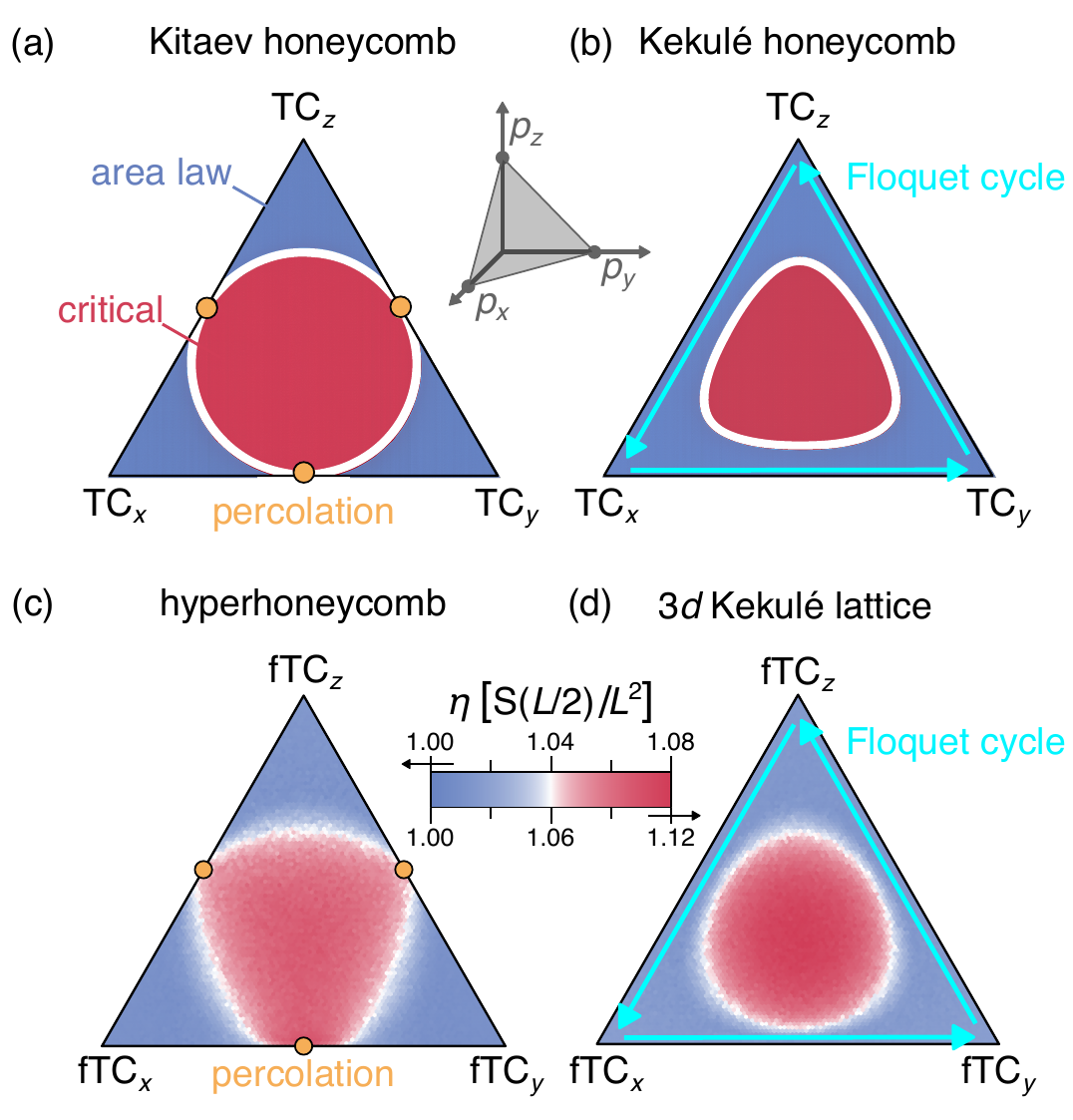}
    \caption{{\bf Random-measurement phase diagrams.}
    (a,b) 2$d$ honeycomb lattices with (a) Kitaev and (b) Kekul\'e edge colorings (adapted from Ref.~\cite{Klocke2025}). (c,d) 3$d$ tricoordinated lattices with (c) the hyperhoneycomb lattice and (d) our lattice. The corners of the phase diagrams correspond to single-color, measurement-dominated limits, realizing a (TC/fTC)-like area-law phase in 2$d$/3$d$ with half-system entanglement scaling as $S(L/2)\propto L^{d-1}$. The central region is critical, with $S(L/2)\propto L^{d-1}\log L$. 
    Note that for (a) and (c) the midpoint of each edge is a $(1+1)$D percolation critical point. The $z\!\rightarrow\!x\!\rightarrow\!y\!\rightarrow\!z$ Floquet cycle can be viewed as a deformation connecting the three corners; in (b) and (d) this path can be chosen to avoid crossing the critical region.
    For the 3$d$ systems we have simulated linear system sizes $L=8,12$ ($L=4, 8$)
    with a total of up to $6,912$ ($12,288$) qubits in total for the hyperhoneycomb and 3$d$ Kekul\'e lattices, respectively. }
    \label{fig:random_measurement}
\end{figure}

Near a color-$c$ corner, where $c$-bonds are measured much more frequently than the other two, the dynamics realizes an extended area-law phase [Fig.~\ref{fig:random_measurement}(a,b)]. This regime can be viewed as a TC-like phase (closely related to the gapped $\mathbb{Z}_2$ topological phase in the Hamiltonian model). With periodic boundary conditions and sufficiently large linear size $L$, the probability that a \emph{nonlocal} logical operator is measured is exponentially suppressed in $L$. By contrast, plaquette syndromes are generated at an $\mathcal{O}(1)$ rate, although the resulting syndrome extraction is less structured than in Floquet measurement schedules. As a consequence, quantum information can, in principle, be encoded and preserved for long times in this area-law regime.

Moving away from the corners, frustration between the three mutually non-commuting measurement types becomes more effective at generating long-range entanglement. In the central region of the phase diagram, the monitored dynamics enters an extended critical regime in which the half-system entanglement entropy scales as
\begin{equation}
S(L/2)\sim L\ln L,
\end{equation}
suggesting an entanglement structure reminiscent of fermionic systems with an extended set of low-energy modes (e.g., a Fermi-surface-like structure, with details depending on the lattice and protocol). In what follows, we use these two regimes -- a TC-like area-law phase near the corners and an enlarged critical region away from them -- as reference points when analyzing the corresponding 3$d$ dynamics and its dependence on lattice geometry.

\subsection{Phase diagram on the 3$\mathbf{d}$ lattices}
Crucially, along an edge of the phase diagram where one measurement type is absent (e.g., $p_z=0$), the dynamics reduces to a two-color monitored Kitaev model. This edge theory can be mapped onto a projective transverse-field Ising model (PTIM)~\cite{Ippoliti2021, Lavasani2023, Sriram2023}, which in turn admits a mapping to classical two-dimensional percolation in spacetime (for the standard PTIM, this corresponds to bond percolation on the square lattice). The percolation critical point lies precisely at the midpoint of the edge, i.e., at equal measurement rates of the two remaining colors. Whether this midpoint corresponds to a \emph{genuine} critical point depends on the geometry of the two-color subgraph obtained by deleting the absent color: if the resulting subgraph decomposes into only finite loops, the dynamics factorizes into finite components and no true criticality can develop in the thermodynamic limit; if it contains non-contractible loops or bi-infinite chains, the percolation transition survives and the edge midpoint is critical. This observation directly connects to the Floquet-code construction: when deleting any one color yields only finite loops, one can design a Floquet cycle that preserves logical information throughout; conversely, the presence of non-contractible structures obstructs such a construction. We therefore obtain a geometric criterion for the presence or absence of criticality along the edges of the phase diagram in three dimensions, which we verify numerically below.

\subsubsection*{Numerical simulations and phase diagrams}
To determine the phase diagram of monitored 3$d$ Kitaev models for the two lattice geometries at hand, we compute the scaling of $S(L/2)$ numerically using Clifford-circuit simulations using the stabilizer tableau formalism (nicely implemented in the package \texttt{QuantumClifford.jl}~\cite{QuantumClifford}). To reduce transients, we initialize the system in a pure state that is a simultaneous eigenstate of all $ZZ$ operators on $z$-bonds, all plaquette operators, and a choice of three independent logical operators. We then perform random measurements according to the prescribed rates up to $t=20$ to obtain the steady state, where time is measured in units of $N$ elementary measurements with $N$ the number of qubits.

To distinguish an area-law phase from the $L^2\log L$ regime, we consider the `two-size ratio'
\begin{equation}
    \eta\!\left[\frac{S(L/2)}{L^{2}}\right]\equiv 
    \frac{S(L_2/2)/L_2^{2}}{S(L_1/2)/L_1^{2}}
\end{equation}
for two distinct linear system sizes $L_2 > L_1$.
In an area-law phase $S(L/2)\propto L^{2}$, this ratio approaches~1, while in a critical regime \(S(L/2)\propto L^{2}\log L\), it exceeds~1. Figure~\ref{fig:random_measurement}(c,d) shows false-color maps of $\eta\!\left[S(L/2)/L^2\right]$ extracted from system sizes $(L_1,L_2)=(8, 12)$ on the hyperhoneycomb lattice and $(L_1,L_2)=(4, 8)$ on our 3$d$ Kekul\'e lattice. In both cases, we observe the same qualitative trend as in 2$d$: area-law phases near the corners (characterized by $\eta\!\left[S(L/2)/L^2\right]\approx 1$) and an extended critical region away from them.
We observe a general tendency that gapless phases of the Hamiltonian Kitaev models evolve into (distinct) critical phases under monitored dynamics, with a critical region that is {\it enlarged} compared to the Hamiltonian case~\footnote{For the Hamiltonian Kitaev model, the (hyper)honeycomb lattice geometries exhibit a well-known triangular-shaped critical region in the center of the phase diagram~\cite{Kitaev2006,OBrien2016}, obtained by connecting the midpoints of the three edges. For the Kekul\'e-Kitaev model, the  Hamiltonian ground-state phase diagram exhibits a singular critical point in the center of the phase diagram~\cite{Schmidt10honeycomb}, while the 3$d$ Kekul\'e-Kitaev model exhibits 
a critical region along a short line-like segment emanating from the isotropic point, see the Appendix~\ref{app:hamiltonian_phase_diagram}.}.
 
Importantly, along the edges of the phase diagram, we find that the hyperhoneycomb lattice exhibits a percolation critical point at the midpoints of all three edges, while our 3$d$ Kekul\'e-Kitaev lattice avoids criticality along the edges. This is consistent with our geometric criterion and, in particular, with the fact that deleting one color from the hyperhoneycomb lattice leaves bi-infinite chains, whereas deleting one color from our lattice leaves only finite loops.
A useful connection between the Floquet code and the random measurement dynamics emerges here. The $z\!\rightarrow\!x\!\rightarrow\!y\!\rightarrow\!z$ Floquet cycle can be viewed as a path connecting the three corners of the measurement diagram -- in our lattice, this path can be chosen to avoid crossing any critical region or point, consistent with the existence of a Floquet code that preserves logical information throughout the cycle.\\

Let us close by stating that while our geometric criterion organizes the random-measurement phase diagram efficiently, it would be valuable to develop an analogue of a ``Majorana band structure'' (or other finer invariant) for $3d$ random measurement dynamics -- potentially requiring effective loop-model descriptions and large-scale numerics to capture distinctions that geometry alone cannot resolve~\cite{Klocke2025}.

\section{Discussion}
\label{sec:discussion}

A key motivation for targeting $3d$ topological codes is the prospect of fault-tolerant non-Clifford logical gates. Different topological orders come with different emergent symmetry structures, and logical gates can often be understood as implementations of these symmetries (e.g., via Wilson/'t~Hooft-type operators)~\cite{Yoshida2015, Beverland2016, Yoshida2017, Webster2018}. 
The 3$d$ fTC, with its robust $(3+1)D$ topological order, is in particular expected to support a richer set of such (emergent) symmetries than its bosonic counterpart, opening the door to intrinsically 3$d$ fault-tolerant non-Clifford logical gates~\cite{Barkeshli2024_2,Barkeshli2024}. Preserving \emph{three} logical qubits throughout the cycle is especially appealing in this context, since prominent non-Clifford candidate gates of the $3d$ fTC, such as the $\mathrm{CCZ}$ gate, naturally act on {\it triples} of logical qubits. At present, however, it remains an open problem to turn these symmetries -- despite the fact that the relevant operator content is naturally formulated in the TQFT language --  into explicit microscopic \emph{local} implementations (finite-depth unitaries and/or measurement protocols) compatible with the Floquet setting.

A separate future direction is suggested by the bosonic case: the color code admits a useful multi-copy viewpoint -- it is locally equivalent to multiple copies of TC order~\cite{Kubica2015}. Recent work on \emph{dynamic automorphism} codes makes this perspective operational by using short local measurement sequences, interpretable as reversible transitions through a suitable parent model, thereby enacting nontrivial code automorphisms and, in $3d$, realizing a non-Clifford logical gate by adaptive measurements~\cite{Davydova2024}. This motivates asking whether multiple copies of the fTC admit an analogous parent description with a transparent automorphism structure, and whether our Floquet sequence (or generalizations thereof) can be understood as a controlled condensation path or a nontrivial automorphism acting across copies. Establishing such a multi-copy framework could provide a systematic route to both gate design and new Floquet constructions.

On the error-correction side, a full fault-tolerance analysis should characterize the effective noise model of the schedule and quantify thresholds and logical error rates~\cite{Gidney2021, Gidney2022}; decoding may naturally separate into a loop-like sector (suggesting single-shot-style redundancy), and a point-defect sector (suggesting matching-type decoders)~\cite{Kulkarni2019, Kubica2019, Vasmer2021}. 
Recent work identifies the fTC as a Hamiltonian model whose low-temperature thermal states {\it remain} long-range entangled, protected by an anomalous 2-form symmetry tied to fermionic Wilson loops~\cite{Zhou2025}; whether this structure has any practical implications for active decoding, especially in the matching-like sector, remains an open question.

It is also tempting to seek a measurement-based quantum computation (MBQC) interpretation of our construction. In two spatial dimensions, circuit-, measurement-, fusion-, and Floquet-based approaches can be related within a single spacetime picture via local equivalence transformations~\cite{Bombin2024}. It would be interesting to ask whether our $3d$ Floquet fTC admits an analogous realization from a $D=4$ spacetime setting, and what this perspective implies for schedule design and logical gates. This connection could provide both conceptual unification and practical tools (e.g., systematic ways to derive schedules and logical gates from a higher-dimensional resource-state perspective). 

Overall, our construction provides a concrete, geometry-driven route to Floquet-generated $3d$ fTC order with persistent logical information. We expect that clarifying the emergent symmetry implementations, possible multi-copy parent structures, and decoding/robustness properties will determine how far this platform can be developed toward fault-tolerant logical control and a sharper understanding of $3d$ measurement-induced phases. \\

{\it Data availability. -- }
The numerical data shown in the figures are available on Zenodo~\cite{zenodo_repository}.

\acknowledgments
We thank Bo Han and Nathanan Tantivasadakarn for insightful discussions.
We acknowledge support, in part, by the Deutsche Forschungsgemeinschaft (DFG, German Research Foundation) under Germany's Excellence Strategy—Cluster of Excellence Matter and Light for Quantum Computing (ML4Q) EXC 2004/1 -- 390534769 as well as within the CRC network TR 183 (Project Grant
No. 277101999) as part of subproject B01.
ST is grateful to the Kavli Institute for Theoretical Physics (KITP), which is supported by the National Science Foundation by grant NSF PHY-2309135, and its summer 2025  program ``Noise-robust Phases of Quantum Matter", where part of this work was carried out.
Our numerical simulations were performed on the Noctua2 cluster at PC$^2$ in Paderborn and the RAMSES cluster at RRZK Cologne.

\appendix

\section{Generating Kekul\'e-type edge colorings from planar graphs}
\label{app:2d_models}

\begin{figure}
    \centering
    \includegraphics[width=1.0\columnwidth]{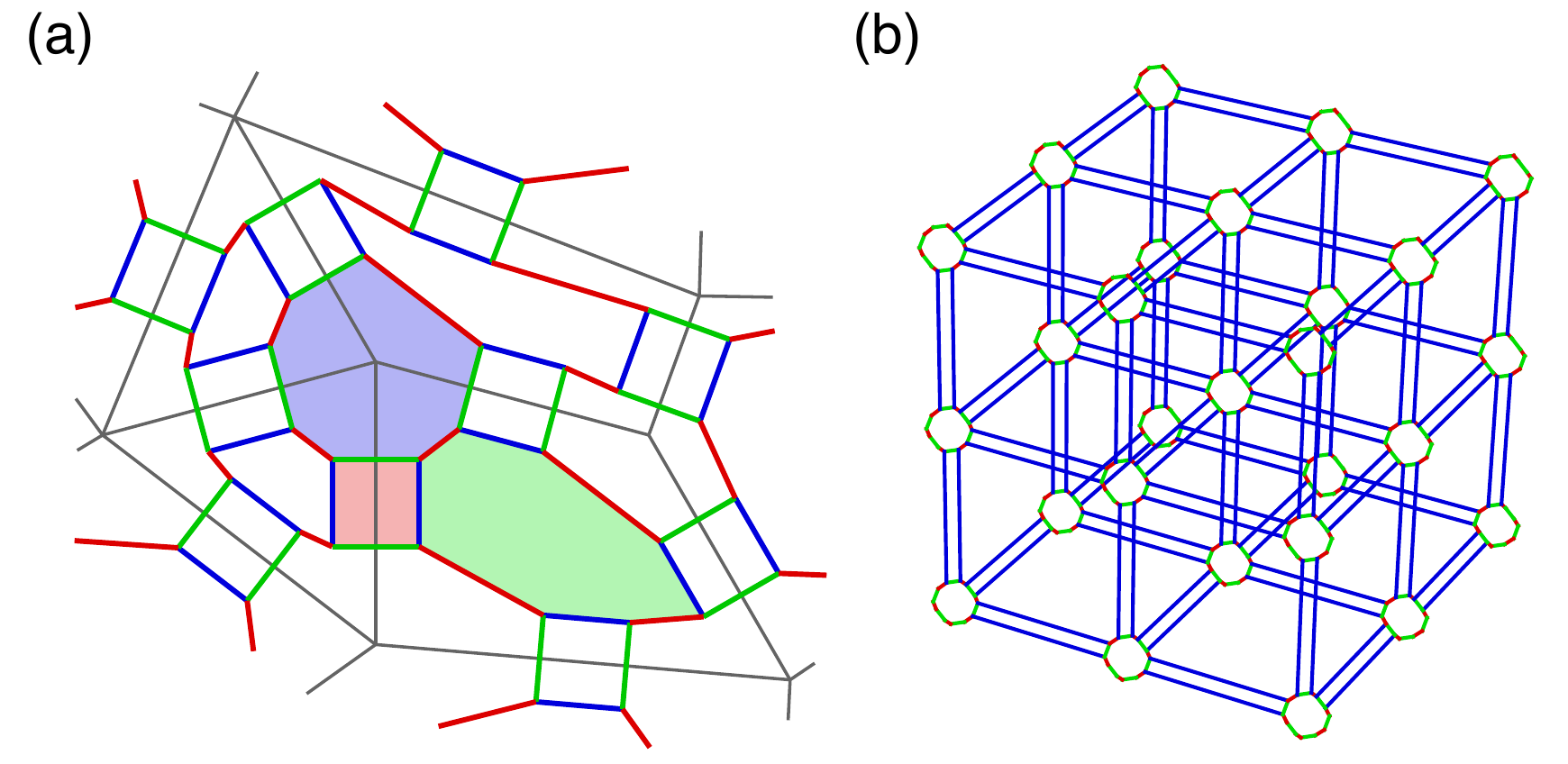}
    \caption{{\bf Generating Kekul\'e-type edge colorings from parent graphs.}
    (a) Starting from a parent planar graph (gray), the decoration described in App.~\ref{app:2d_models} produces a tricoordinated lattice with a proper three-edge coloring and only two-colored plaquettes.
    (b) Applying the same decoration to the cubic lattice yields a tricoordinated, properly three-edge-colored 3$d$ graph, but it fails to satisfy the finite-loop property for all colors; in the example shown, deleting all $y$ bonds leaves bi-infinite chains.}
    \label{fig:appendix_a}
\end{figure}

There are infinitely many 2$d$ tricoordinated lattices satisfying the design conditions (1)--(3) discussed in the main text. An explicit family can be generated by decorating an arbitrary \emph{parent} planar graph $H$ with vertex set $V(H)$, edge set $E(H)$, and face set $F(H)$, as illustrated in Fig.~\ref{fig:appendix_a}(a).
For each edge $e=(u,v)\in E(H)$, replace $e$ by a four-cycle $C_{yz}(e)$ whose edges alternate in colors $y$ and $z$, arranged so that the two $z$ edges run parallel to $e$ and the two $y$ edges sit at the two ends. Next, add $x$-colored edges to ``enclose'' each parent vertex: for each $v\in V(H)$, collect the $2\deg(v)$ vertices contributed by the incident four-cycles on the $v$-side and connect them in cyclic order (as fixed by the planar embedding) by $x$ edges, forming an $x$--$y$ loop around $v$. After all such $x$--$y$ loops are added, each face $f\in F(H)$ of the planar embedding is bounded by an $x$--$z$ cycle in the decorated graph $G$.

The resulting planar graph $G$ is tricoordinated and properly three-edge-colored. Deleting any one color leaves only a disjoint union of finite loops: removing $x$ isolates the $y$--$z$ four-cycles $C_{yz}(e)$, removing $z$ leaves the $x$--$y$ loops around vertices, and removing $y$ leaves the $x$--$z$ loops around faces [see Fig.~\ref{fig:appendix_a}(a)]. Since the decoration is local once the cyclic order at each vertex is specified, the construction applies to arbitrary planar embeddings (including, e.g., hyperbolic tilings). As an example, applying this procedure to the square lattice produces the square--octagon lattice.

Planarity enters in two distinct ways. First, the planar embedding fixes a cyclic order of edges around each parent vertex, which makes the choice of the enclosing $x$-colored loop canonical. In three dimensions one can still connect the local decorations near a vertex to form such loops, but the choice is no longer unique. More importantly, in 2$d$ the embedding supplies a canonical set of faces with contractible boundary cycles, so the remaining $x$--$z$ edges are forced to close into short, local plaquettes of $\mathcal{O}(1)$ length. By contrast, in three dimensions an embedded \emph{graph} does not, by itself, induce an analogous notion of faces. Consequently, the leftover two-color subgraph (e.g., the $x$--$z$ edges) is not forced to decompose into short, contractible cycles; it can instead contain extended cycles that do not bound any local plaquette and may be homologically nontrivial. Figure~\ref{fig:appendix_a}(b) illustrates this obstruction for a cubic-lattice attempt, where deleting $y$ leaves bi-infinite chains.

Even in such partially successful 3$d$ constructions, the finite-loop condition (2) can still hold for \emph{two} of the three colors (here, $x$ and $z$). Therefore, one could attempt schedules in which $x$ and $z$ rounds are never adjacent (so that the protocol avoids the problematic $x\leftrightarrow z$ transition), supplemented by additional rounds to access the remaining plaquette syndromes, as in the 3$d$ construction in the main text. While this is less uniform than the 3$d$ Kekul\'e--Kitaev lattice introduced there, the parent-graph structure may still provide a useful starting point for systematic schedule design. We leave a detailed exploration of such parent-graph-based 3$d$ schedules for future work.

\section{Hamiltonian phase diagram of the 3$\mathbf{d}$ Kekul\'e--Kitaev model}
\label{app:hamiltonian_phase_diagram}

The Hamiltonian of the 3$d$ Kekul\'e--Kitaev model has the standard form
\begin{equation}
    H=\sum_{\langle i,j\rangle} J_{\alpha_{ij}} \sigma_i^{\alpha_{ij}} \sigma_j^{\alpha_{ij}}\,,
\end{equation}
where $\langle i,j\rangle$ denotes nearest-neighbor pairs of lattice sites, $\alpha_{ij}\in\{x,y,z\}$ is the bond color of the edge connecting sites $i$ and $j$, and $J_{\alpha_{ij}}$ is the corresponding coupling strength. As in other Kitaev-type models, this Hamiltonian can be mapped to free Majorana fermions coupled to a static $\mathbb{Z}_2$ gauge field~\cite{Kitaev2006}. The gauge field is specified by link variables $u_{ij}=\pm 1$, and the gauge-invariant flux through an elementary plaquette $p$ is
\begin{equation}
    W_p=\prod_{(i,j)\in\partial p} u_{ij}\,,
\end{equation}
where the product runs along the boundary $\partial p$.

In Kitaev-type models, the energetically preferred flux sector can typically be identified using Lieb's flux-phase theorem~\cite{Lieb1994} (beyond the exact applicability of the theorem as demonstrated for a variety of 3$d$ lattice geometries \cite{Eschmann2020}). For a bipartite hopping Hamiltonian at half filling with a reflection symmetry that leaves a plaquette invariant (i.e., maps its boundary cycle to itself and does not pass through any lattice site), the flux through that plaquette is pinned to either $0$ or $\pi$; for an even-length plaquette of perimeter $|p|$, the energy-minimizing choice is
\begin{equation}
    W_p =
    \begin{cases}
        +1, & |p| = 4n+2 \quad (0\text{-flux}),\\
        -1, & |p| = 4n \quad (\pi\text{-flux}).
    \end{cases}
    \label{eq:lieb_flux_rule}
\end{equation}
Following earlier work on 3$d$ Kitaev models~\cite{OBrien2016,Eschmann2020}, we adopt these flux assignments~\eqref{eq:lieb_flux_rule} also for our 3$d$ Kekul\'e lattice, in order to determine the ground-state phase diagram. For every elementary plaquette type except $p_3$, there exists a site-avoiding mirror plane that leaves the plaquette invariant, so Lieb's theorem fixes the corresponding flux. The $p_3$ plaquette does not admit such a mirror plane; nevertheless, we choose the same assignment, motivated by the $J_z=0$ limit in which $p_3$ reduces to a length-12 Kitaev chain whose lowest-energy sector corresponds to $\pi$ flux, and we assume this assignment persists throughout the phase diagram.

\begin{figure}
    \centering
    \includegraphics[width=1.0\columnwidth]{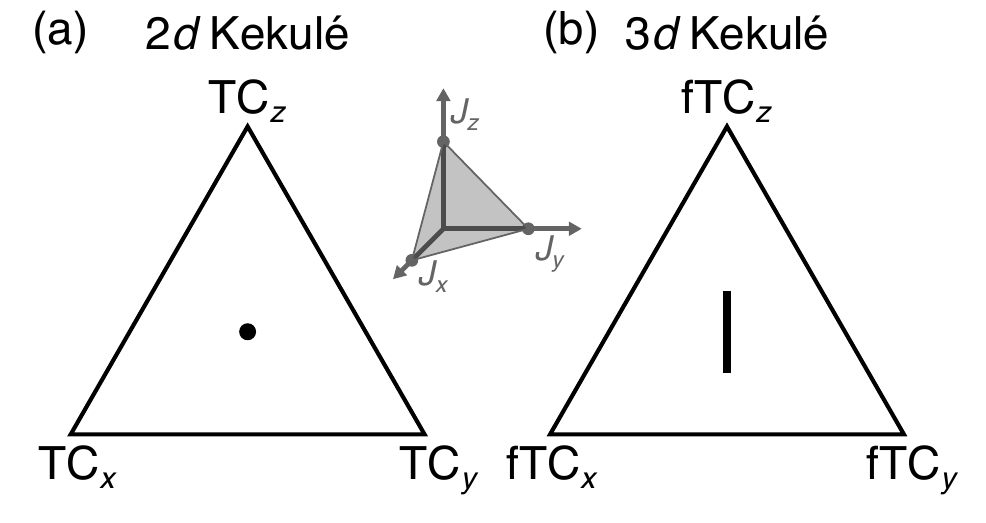}
    \caption{{\bf Hamiltonian phase diagram.}
    (a) 2$d$ honeycomb lattice with Kekul\'e edge coloring. The gapless region is a single point at the isotropic coupling $J_x=J_y=J_z$. (b) 3$d$ Kekul\'e--Kitaev lattice. The gapless region is a short line-like segment emanating from the isotropic point along the $J_x=J_y$ direction, approximately spanning $J_z \in [0.239,\,0.411]$ at our $\mathbf{k}$-grid resolution.
    }
    \label{fig:Hamiltonian_phase_diagram}
\end{figure}

A convenient translation-invariant representative of this flux sector is specified by the link variables $u_{ij}$ within the 24-site unit cell shown in Fig.~\ref{fig:lattice}(c). Using the convention that we tabulate $u_{ij}$ for an oriented bond $(i,j)$ with $i<j$, we choose
\begin{equation}
u_{ij}=
\begin{cases}
-1, & (i,j)\in\mathcal B,\\
+1, & \text{otherwise},
\end{cases}
\qquad\text{and}\qquad
u_{ji}=-u_{ij},
\label{eq:reference_gauge}
\end{equation}
with
\begin{equation}
    \begin{aligned}
        \mathcal B=\{&(1,8),(2,9),(3,11),(6,7),(6,10),(8,12),\\
        &(13,20),(14,21),(15,23),(18,19),(18,22),(20,24),\\
        &(9,21),(11,23)\}.
    \end{aligned}
\end{equation}

With the static gauge field configuration fixed by Eq.~\eqref{eq:reference_gauge}, the Majorana hopping Hamiltonian reads
\begin{equation}
    H=i\sum_{\langle i,j\rangle} J_{\alpha_{ij}} u_{ij} c_i c_j \,.
    \label{eq:majorana_hopping}
\end{equation}
The edge coloring doubles the translation period along both $\vec a_1$ and $\vec a_2$, so the Majorana hopping problem has a 96-site unit cell (in the unrotated convention of Fig.~\ref{fig:lattice}). With periodic boundary conditions, Eq.~\eqref{eq:majorana_hopping} can be diagonalized in momentum space, yielding the single-particle Majorana spectrum $\varepsilon_n(\mathbf{k})$ ($n=1,\dots,96$).

We numerically evaluate the bulk Majorana gap
\begin{equation}
    \Delta \equiv \min_{\mathbf{k}, n}\, \abs{\varepsilon_n(\mathbf{k})}\,,
\end{equation}
by sampling a uniform $\mathbf{k}$-grid over the Brillouin zone. The resulting Hamiltonian phase diagram is shown in Fig~\ref{fig:Hamiltonian_phase_diagram}(b). It contains a short gapless segment emanating from the isotropic point $J_x=J_y=J_z$ along the $J_x=J_y$ direction; at our $\mathbf{k}$-grid resolution it approximately spans $J_z \in [0.239,\,0.411]$ and terminates before reaching the boundary of the phase diagram. The remainder of the parameter space is fully gapped. Along the edges (where one coupling vanishes), the Majorana hopping problem reduces to decoupled one-dimensional subsystems equivalent to a Kitaev chain~\cite{Kitaev2001,Brzezicki2007,Feng2007} of \emph{fixed length}, and therefore exhibits a finite-size gap even at parameter values that would be critical in the infinite-chain limit. This is consistent with the absence of criticality along the edges of the random-measurement phase diagram discussed in the main text.


\clearpage

\bibliographystyle{plainnat}
\bibliography{ref}

\end{document}